\documentclass[12pt]{article}
\usepackage{geometry}
\usepackage{subfigure}
\makeatletter
\newcommand\ackname{Acknowledgements}
\if@titlepage
  \newenvironment{acknowledgements}{
      \titlepage
     \null\vfil
     \@beginparpenalty\@lowpenalty
      \begin{center}
        \bfseries \ackname
        \@endparpenalty\@M1
      \end{center}}
     {\par\vfil\null\endtitlepage}
\else
  
\usepackage{mathrsfs}
\usepackage{eucal}
\usepackage{amssymb,amsmath}

\usepackage{pgf}
\usepackage{tikz}
\usepackage{cite}
\usepackage{graphicx}
\usepackage{color}
\usepackage{amssymb}
\usepgflibrary{arrows}
\usetikzlibrary{calc,angles,positioning,intersections,quotes,backgrounds,decorations.pathreplacing}
\usetikzlibrary{decorations.markings}
\usepackage{amsmath}
\usepackage{amsthm}
\usepackage{accents}

\theoremstyle{remark}
\theoremstyle{definition}

\newcommand{\nn}{\nonumber}
\usepackage{comment}
\begin{document}
\title{\textbf{Geodesic compatibility: goldfish systems }}
\author{Worapat Piensuk$^{\dagger}$ and  Sikarin Yoo-Kong$^{\dagger\dagger } $ \\
\small $^\dagger$ \emph{Department of Mechanical Engineering, Faculty of Engineering, }\\
 \small \emph{King Mongkut's University of Technology Thonburi, Thailand, 10140}\\
\small $^{\dagger\dagger} $ \emph{The Institute for Fundamental Study (IF),} \\ \small\emph{Naresuan University, Phitsanulok, Thailand, 65000.}\\
}
\date{}
\maketitle

\abstract
To capture a multidimensional consistency feature of integrable systems in terms of the geometry, we give a condition called \emph{geodesic compatibility} implying the existence of integrals in involution of the geodesic flow. The geodesic compatibility condition is constructed from a concrete example namely the integrable Calogero's goldfish system through the Poisson structure and the variational principle. The geometrical view of the geodesic compatibility gives compatible parallel transports between two different Hamiltonian vector fields.

\section{Introduction}
In the recent years, the multidimensional consistency feature of integrable systems has been extensively studied by many people in the field. This intriguing feature first arose in the level of discrete integrable systems, namely, the \emph{consistency around the cube} (CAC) \cite{F20011,F20012,F20021,FSU1} such that there exists a set of compatible equations defined in each subspace corresponding to the number of independent variables. This means that it allows us to consistently embed the difference equations in a multidimensional discrete space. In the context of Hamiltonian systems, the Liouville integrability is a natural criterion to test the system in question \cite{ARNO}. The important feature is called the \emph{Hamiltonian commuting flows} which can actually be considered as the multidimensional consistency in the level of the Poisson structure. Such consistency can also be captured in the Lagrangian description known as the Lagrangain multi-forms \cite{SF1}. The main feature for the integrability in this context is called the \emph{closure relation} which implies the existence of infinite paths on the space of independent variables corresponding to a single path on the space of dependent variables with a critical action. 
\\
\\
The Calogero-Moser (CM) type systems, Ruijsanaars-Schneider (RS) type systems and Calogero's goldfish (GF) type systems are well-known integrable one-dimensional many-body systems \cite{CM,RS,GFCal} in the context of Liouville integrability. Furthermore, their integrability can also be exhibited through the Lagrangian 1-form structure \cite{Sikarin1,Sikarin2,Umpon,Umpon1,LAMS4,LAMS5,LAMS3,Chis}. Intriguingly, for the GF systems, Hamiltonians are all written with exponential of conjugate momenta, and their equations of motion are perfectly in a form of geodesic representation. The geodesic interpretation of GF models was first investigated in \cite{Jens1}, and it was found that the Riemann curvature tensor for the case of rational GF model vanishes suggesting that the evolution of this system is indeed a free geodesic motion in Cartesian-like coordinates under the coordinate transformation. 
\\
\\
%The geometrical condition for geodesic compatibility of the systems with pseudo-Riemannian metrics has been proposed by Topalov \cite{Topa1}. A pair of geodesic flows is said to be compatible if their Christoffel symbols satisfy the so-called \textit{$PQ^\epsilon$-projectivity} which is the generalization of geodesic equivalence and holomorphically projective (h-projectivity). If two pseudo-Riemannian metrics obey such condition, it was shown in the paper that the hierarchy of compatible metrics admitting functionally-independent integrals of motion which are in involution with each other can be produced. An example for the construction has also been made for the case of complex projective plane \textbf{CP}$^{n}$.
As we mentioned earlier that the GF models are the integrable systems exhibiting the multidimensional consistency through the Hamiltonian commuting flows and the closure relation and, since GF models are also integrable geodesic flows, in the present paper, we would like to capture the multidimensional consistency from the geometrical point of view through the \textit{general} metric tensors. In section \ref{section2}, we provide a brief review on the geodesic flows and a criterion for their integrability. The GF systems are also presented together with their geodesic interpretation. In section \ref{section3}, the condition on metric tensors called the \emph{geodesic compatibility}\footnote{The geometrical condition for geodesic compatibility of the systems with pseudo-Riemannian metrics has been investigated by Topalov \cite{Topa1}. A pair of geodesic flows is said to be compatible if their Christoffel symbols satisfy the so-called \textit{$PQ^\epsilon$-projectivity}. If two pseudo-Riemannian metrics obey such condition, the hierarchy of compatible metrics admitting functionally-independent integrals of motion, which are in involution with each other, can be produced.} is derived from the commuting Hamiltonian flows and the variational principle on the space of time variables. Both rational and hyperbolic GF models are explicitly used to verify the condition. In section \ref{section4}, the interpretation of the geodesic compatibility condition in the geometrical point of view is presented. In section \ref{section5}, the summary, as well as remarks, is given .

\section{Preliminaries}\label{section2}
%%%%%%%%%%%%%%%%%%%%%%%%%%
\subsection{Integrable geodesic flows}
\par Suppose there is an $N$-dimensional manifold $\CMcal M$ equipped with the metric tensor $g(q)$, where $q=(q^1,q^2,...,q^N)$ is a set of local coordinates, and a pair $(\CMcal M,g)$ forms a well-known (smooth) Riemannian manifold. Let a smooth curve $\gamma(q(t),\frac{d}{dt}q(t))$ be a geodesic defined on the tangent bundle $T\CMcal M$. Mathematically, a geodesic flow is a family of the diffeomorphisms $\phi_t$ of the tangent bundle such that each point on the geodesic can be expressed as \cite{Patgeo}
\begin{eqnarray}
\phi_t \left(q(0),\frac{d}{dt}{q}(0)\right) := \left(q(t),\frac{d}{dt}q(t)\right)\;.\label{Map1}
\end{eqnarray}
Let us now define $S\CMcal M$ as a unit tangent bundle which is a subset of $T\CMcal M$ where $dq/dt$ has a unity norm. We find that $S\CMcal M$ is preserved under the map defined in \eqref{Map1} along the curve $\gamma$, i.e., for any $(q, dq/dt) \in S\CMcal M$, $\phi_t (q, dq/dt) \in S\CMcal M$. 
\\
\\
In Hamiltonian context, the geodesic flow is the trajectory that describes the evolution for a system of equations
\begin{eqnarray}\label{HE}
\frac{dq^i}{dt}=\frac{\partial \mathscr{H}}{\partial p_i}\;,\;\;\;-\frac{dp_i}{dt}=\frac{\partial \mathscr{H}}{\partial q^i}\;,\;\;i=1,2,...,N\;,
\end{eqnarray}
on the cotangent bundle $T^*\CMcal M$. The Hamiltonian $\mathscr{H}(q,p)$ is given in the form
\begin{eqnarray}\label{Hge}
\mathscr{H} (q,p)=\frac{1}{2} \sum_{i,j} ^Ng^{ij}(q)p_i  p_j = \frac{1}{2} g^{ij}p_i  p_j\;,
\end{eqnarray}
where $(q,p)\equiv(q^1, q^2, ..., q^N, p_1, p_2, ..., p_N)$ are the canonical coordinates on a $2N$-dimensional phase space, and $g^{ij}$ are the elements in the metric tensor such that $p_j=g_{ij}\frac{d}{dt}q^i$. With a given Hamiltonian \eqref{Hge}, equation \eqref{HE} reads
\begin{eqnarray}\label{HE2}
\frac{dq^i}{dt}=g^{ij}p_j\;,\;\;\;-\frac{dp_i}{dt}=\frac{1}{2}\frac{\partial g^{jk}}{\partial q^i}p_jp_k\;
\end{eqnarray}
resulting in the geodesic equations 
\begin{eqnarray}\label{GE}
\frac{d^2q^i}{dt^2}+\Gamma^i_{jk}\frac{dq^j}{dt}\frac{dq^k}{dt}=0\;,
\end{eqnarray}
where $\Gamma^i_{jk}$ are the affine connection given by
\begin{equation}
\Gamma^i_{jk}=\frac{1}{2}g^{im}\left(\frac{\partial g_{jm}}{\partial q^k}+\frac{\partial g_{km}}{\partial q^j}-\frac{\partial g_{jk}}{\partial q^m} \right)\;.
\end{equation}
 %and the Einstein summation convention, which will be used throughout this report when there exists one index that appears twice as lower and upper indices, is assumed.
In general, the geodesic on a closed Riemannian manifold can be globally complicated, but still regular, and of course not chaotic. Global regular behaviour is a main characteristic property of integrable geodesic flows which are defined as follows. 
\\
\\
\textbf{Definition}: The geodesic flow is said to be \textit{completely Liouville integrable} if there exists a set of $N$ functions defined on the phase space $\{ F_1(q,p), F_2(q,p), ..., F_N(q,p)\}$ which satisfies the following requirements:
\begin{itemize}
\item They are integrals of the geodesic flow, i.e., constant along each geodesic line.
\item They are commuting with respect to Poisson bracket on $T^*\CMcal M$, i.e., $\{F_i, F_j\}=0$, where $i\neq j=1,2,...,N$
\item They are functionally independent on $T^*\CMcal M$. In other words, the gradients of every integrals are linearly independent.
\end{itemize}
%\end{definition}
We find that it is not difficult to obtain the Lagrangian associated with the Hamiltonian \eqref{Hge} 
\begin{eqnarray}\label{Lge}
\mathscr{L}\left(q,\frac{dq}{dt}\right) =\frac{1}{2}g_{ij}\frac{d q^i}{dt}  \frac{d q^j}{dt}\;,
\end{eqnarray}
and the Euler-Lagrange equations
\begin{eqnarray}\label{Lge}
\frac{\partial\mathscr{L}}{\partial q^i}-\frac{d}{dt}\frac{\partial\mathscr{L}}{\partial (\frac{dq^i}{dt})}=0\;
\end{eqnarray}
give us again \eqref{GE}.\\\\
We end this section with some well known examples of topological objects admitting integrable geodesic flows. The 2-Sphere $S^2 := \{(q^1)^2+(q^2)^2+(q^3)^2=1\}$, whose geodesics are equators (the curves on the great circles), and the torus with a flat metric $ds^2 = (dq^1)^2+(dq^2)^2$, whose angle coordinates $\theta_i(t)$ defining the surface is quasi-periodic, i.e., $\theta_i(t) = c_i t$ of period $2\pi$, where $i=1,2$, are classical examples of two-dimensional surfaces with integrable geodesic flows. However, surfaces of revolution admitting non-trivial linear constants of motion called \textit{Clairaut integrals} and surfaces with Liouville metrics $ds^2 = (f(x)+g(x))((dq^1)^2+(dq^2)^2)$ admitting non-trivial quadratic integrals are also examples. \cite{Bol1}

%%%%%%%%%%%%%%%%%%%%%%%%%%

\subsection{GF models as geodesic Hamiltonian flow}
The GF models are the Hamiltonian system \cite{GFCal,GFSur} whose Hamiltonian is given by
\begin{eqnarray}
\mathscr{H}(q,p)=\sum_{i=1}^N e^{ap_i} \prod\limits_{\mathop {j = 1}\limits_{j \ne i}}^N f(q^i-q^j)\;,\label{HGF}
\end{eqnarray}
where $a$ is a parameter, and
\begin{eqnarray}
f(q)=\begin{cases}
      \frac{1}{q} & \text{: rational case}\\
      \frac{1}{\sinh(q)} & \text{: hyperbolic case}\;,
    \end{cases}\nn
\end{eqnarray}
respectively. Using Hamilton's equations, the equations of motion are given by
\begin{eqnarray}
\frac{d^2{q}^i}{dt^2} = \sum_{j\neq i}^{N} \frac{d{q}^i}{dt} \frac{d{q}^j}{dt} W(q^i-q^j) \quad \text{for} \quad i = 1, 2, ..., N\;, \label{eq22}
\end{eqnarray}
where
\begin{eqnarray}
W(q)=\begin{cases}
      \frac{2}{q} & \text{: rational case}\\
      2\coth(q) & \text{: hyperbolic case\;,}
    \end{cases}\nn
\end{eqnarray}
and $\gamma$ is an arbitrary parameter. It accidentally turns out that (\ref{eq22}) are in the form of geodesic equations with the affine connection given by \cite{Jens1}
\begin{eqnarray}
\Gamma_{jk}^i = \delta_{j}^i w_{ik}+\delta_{k}^i w_{ij}\;,\;\;\mbox{and}\;\; w_{ij} =- \frac{1}{2}(1-\delta_{ij})W(q^i-q^j) \;.
\end{eqnarray}
In the rational case, it has been shown that all components of the curvature tensor (Riemann tensor)
\begin{equation}
R^i_{jkl}=\frac{\partial \Gamma^i_{lj}}{\partial q^k}-\frac{\partial \Gamma^i_{kj}}{\partial q^l}+\Gamma^i_{km}\Gamma^m_{lj}-\Gamma^i_{lm}\Gamma^m_{kj}\;
\end{equation}
vanish identically. This means that the evolution of rational GF model is indeed free geodesic, and there exists the Cartesian-like coordinates
\begin{equation} 
x_n[q]=\frac{1}{n!}\sum_{(i_1,i_2,...,i_N)'} q^{i_1}q^{i_2}...q^{i_n}\;, \;n=1,2,..,N\label{XX}
\end{equation}
where$\;'\;$ indicates that the all indices are different, such that the goldfish equations \eqref{eq22} becomes $d^2x_n/dt^2=0$.
\\
\\
The geodesic interpretation for the RS systems and Toda systems had been investigated further. In the case of rational RS system, there is the same structure as the rational GF system. In the hyperbolic RS and relativistic Toda systems, it turns out to be that they are linked to non-metric connections \cite{Gal1,Gal2}. For non-metric case, an investigation on a sufficient and necessary condition for a system with two-dimensional affine connection admitting linear first integrals was given in \cite{Lev1}. The restriction on the form of the affine connection of the Dubrovin–Novikov Hamiltonian formulation of the one-dimensional hydrodynamic system is also analyzed in \cite{Con1}.
%\textcolor{blue}{On the contrary, It had been shown that the curvature tensor for hyperbolic goldfish model does not vanish. In addition, the investigation on both rational and hyperbolic Ruijsenaars-Schneider systems has also been made in \cite{Gal1}. The result from the paper showed similarity to the goldfish models in such a way that the curvature tensor of rational case vanishes, but non-trivial, for hyperbolic case.}
%We leave this section with a point that actually the rational Ruijsenaars-Schneider system 
%\begin{equation}
%\mathscr{H}=
%%\end{equaton}
%is also integrable geodesic flow and the curvature tensor
%%%%%%
%%%%%%
%%%%%%
%%%%%%
%%%%%%
%%%%%%
%%%%%%
%%%%%%
%%%%%%
%%%%%%%%%%%%%%%%%%%%%%%%%%%%
\section{Compatible geodesic flows}\label{section3}
It is known that the GF models \eqref{HGF} are completely integrable \cite{GFSur} and certainly the systems possess the Hamiltonian hierarchies. The first three Hamiltonians of the GF system are  
\begin{eqnarray}
\mathscr{H}=\mathscr{H}_{1} = g^{ij}_1\pi_{i}\pi_{j} \;, \quad \mathscr{H}_{2} = {g}^{ij}_2\pi_{i}\pi_{j} \;, \quad \mathscr{H}_{3} = {g}^{ij}_3\pi_{i}\pi_{j}\;,\;\;i,j=1,2,...,N\;,\label{eq11}
\end{eqnarray}
where $\pi_{i} \equiv e^{p_{i}/2}$ and $p_i$ is the conjugate momentum in canonical coordinates $(p,q)$. The first three metric tensors for the rational case are given by
\begin{eqnarray}
g^{ij}_1 = \delta_{ij}\frac{1}{\prod\limits_{a \neq i}^{N}(q^i-q^a)} \;, \quad
{g}^{ij}_2= \delta_{ij}\frac{\sum\limits_{b\neq i}^{N}q^b}{\prod\limits_{a \neq i}^{N}(q^i-q^a)}\; , \quad
{g}^{ij}_3 = \delta_{ij}\frac{\sum\limits_{n\neq m}^{N}q^m q^n}{\prod\limits_{a\neq i}^{N}(q^i-q^a)}\;\;,
\end{eqnarray}
and the first three metric tensors for the hyperbolic case are given by
\begin{eqnarray}
g^{ij}_1 = \delta_{ij}\frac{1}{\prod\limits_{a \neq i}^{N}\sinh{(q^i-q^a)}} \;, \quad
{g}^{ij}_2= \delta_{ij}\frac{\sum\limits_{b\neq i}^{N}e^{-2q^b}}{\prod\limits_{a \neq i}^{N}\sinh{(q^i-q^a)}}\; , \quad
{g}^{ij}_3 = \delta_{ij}\frac{\sum\limits_{n\neq m}^{N}e^{-2q^m-2q^n}}{\prod\limits_{a\neq i}^{N}\sinh{(q^i-q^a)}}\;\;.
\end{eqnarray}
It is well-known that the Hamiltonian is a time generator and here we definitely have different time variables for each Hamiltonian. The geodesic equations for the Hamiltonian $\mathscr{H}_k$ are given by
\begin{equation}
\frac{d^2q^i}{dt_{1}dt_{k}} = \sum_{j=1,j\neq i}^{N} \frac{d{q}^i}{dt_1} \frac{d{q}^j}{dt_k} W(q^i-q^j) \quad \text{for} \quad k = 1, 2, ..., N\;. \label{eqtk}
\end{equation}
However, we would like to point that the Hamiltonians in \eqref{eq11} are in the pseudo-geodesic form. What we mean by ``pseudo" is that actually this Hamiltonian hierarchy is not explicitly in the form given in \eqref{Hge} since the momenta $\pi_i$ are not canonical variables and of course the Poisson bracket $\{\pi_i,q^j \}\neq \delta_i^j$.
\\
\\
In this section, we are interested in constructing the relation that implies integrability through the structure of the $g$ metric tensors. We first set out to derive the condition directly from the involution of the Hamiltonians, and then we look for the condition from a different perspective, namely, from the variational principle.
\\
\\
\textbf{The Poisson structure}: Given two arbitrary Hamiltonians in the hierarchy 
\begin{eqnarray}
\mathscr{H}_{l} = g_l^{ij}\pi_{i}\pi_{j} \quad \mbox{and} \quad \mathscr{H}_{s} = g_s^{ij}\pi_{i}\pi_{j}\;,\label{eq1}
\end{eqnarray}
the Poisson bracket between them gives 
\begin{eqnarray}
\{\mathscr{H}_{l}, \mathscr{H}_{s} \}&=& \frac{\partial \mathscr{H}_{l}}{\partial q^m}\frac{\partial \mathscr{H}_{s}}{\partial p_m} -\frac{\partial \mathscr{H}_{s}}{\partial q^m} \frac{\partial \mathscr{H}_{l}}{\partial p_m}   \nn\\
&=& \left(\frac{\partial g_l^{ij}}{\partial q^m}\pi_{i}\pi_{j}\right)(g_s^{nk}\pi_{k}(\frac{1}{2}\pi_{n}\delta_{nm})+g_s^{nk}\pi_{n}(\frac{1}{2}\pi_{k}\delta_{km}))\nn\\
&&-\left(\frac{\partial g_s^{ij}}{\partial q^m}\pi_{i}\pi_{j}\right)(g_l^{nk}\pi_{k}(\frac{1}{2}\pi_{n}\delta_{nm})+g_l^{nk}\pi_{n}(\frac{1}{2}\pi_{k}\delta_{km}))\nn\\
&=& \left(\frac{\partial g_l^{ij}}{\partial q^m}\pi_{i}\pi_{j}\right)(\frac{1}{2}g_s^{mk}\pi_{k}\pi_{m}+\frac{1}{2}g_s^{nm}\pi_{n}\pi_{m})\nn\\
&&-\left(\frac{\partial g_s^{ij}}{\partial q^m}\pi_{i}\pi_{j}\right)(\frac{1}{2}g_l^{mk}\pi_{k}\pi_{m}+\frac{1}{2}g_l^{nm}\pi_{n}\pi_{m})\nn\\
&=&\left(\frac{\partial g_l^{ij}}{\partial q^k}g_s^{nk}-\frac{\partial g_s^{ij}}{\partial q^k}g_l^{nk}\right)\pi_{i}\pi_{j}\pi_{n}\pi_{k}\;.\nn
\end{eqnarray}
The involution condition gives
\begin{eqnarray}
\sum_{i,j,n,k = 1}^{N}\left(\frac{\partial g_l^{ij}}{\partial q^k}g_s^{nk}-\frac{\partial g_s^{ij}}{\partial q^k}g_l^{nk}\right)\pi_{i}\pi_{j}\pi_{n}\pi_{k} = 0\;.\label{eq2}
\end{eqnarray}
This equation can be called as the \emph{geodesic compatibility} and can be used as an integrability criterion for the pseudo-geodesic Hamiltonian systems.
\\
\\
\textbf{The variational principle}: The action functional of a system with $N$ independent variables is given by 
\begin{eqnarray}
S &=&\int_{\mathscr{C}} \sum_{k=1}^N(p_i q^i_{t_k} - g_k^{ij}\pi_{i}\pi_{j})dt_k\;,
\end{eqnarray}
where $\mathscr{C}$ is a curve on the space of independent variables and $q_{t_k} \equiv \partial q/\partial t_k$. Now we introduce a time-parameterised variable $s$, $s_0\leq s\leq s_1$, such that
\begin{eqnarray}
S &=& \int_{s_0}^{s_1}\bigg[ \sum_{k=1}^N(p_i{q}^i_{t_k} - g_k^{ij}\pi_{i}\pi_{j})\frac{dt_k}{ds}\bigg]ds \;.\nn
\end{eqnarray}
The variation of action according to the local deformation on $t_l$-$t_s$ plane ($l\neq s$) is 
\begin{eqnarray}
\delta S &=& \int\bigg\{\bigg[p_iq_{t_l}^i\frac{d\delta t_l}{ds}+\left(\frac{\partial p_i}{\partial t_l}q_{t_l}^i\delta t_l+\frac{\partial p_i}{\partial t_2}q_{t_l}^i\delta t_s+p_i\frac{\partial q_{t_l}^i}{\partial t_l}\delta t_1+p_i\frac{\partial q_{t_l}^i}{\partial t_s}\delta t_s-\frac{\partial g_l^{ij}}{\partial t_l}\pi_{i}\pi_{j}\delta t_l-\frac{\partial g_l^{ij}}{\partial t_s}\pi_{i}\pi_{j}\delta t_s\right.\nn\\
&-&\left.\frac{1}{2}g_l^{ij}\frac{\partial p_i}{\partial t_l}\pi_{i}\pi_{j}\delta t_l-\frac{1}{2}g_l^{ij}\frac{\partial p_i}{\partial t_s}\pi_{i}\pi_{j}\delta t_s-\frac{1}{2}g_l^{ij}\frac{\partial p_j}{\partial t_l}\pi_{i}\pi_{j}\delta t_l-\frac{1}{2}g_l^{ij}\frac{\partial p_j}{\partial t_s}\pi_{i}\pi_{j}\delta t_s\right)\frac{dt_1}{ds}-g_l^{ij}\pi_{i}\pi_{j}\frac{d\delta t_1}{ds}\bigg]\nn\\
&+&\bigg[p_iq_{t_s}^i\frac{d\delta t_s}{ds}+\left(\frac{\partial p_i}{\partial t_l}q_{t_s}^i\delta t_l+\frac{\partial p_i}{\partial t_s}q_{t_s}^i\delta t_s+p_i\frac{\partial q_{t_s}^i}{\partial t_l}\delta t_l+p_i\frac{\partial q_{t_s}^i}{\partial t_s}\delta t_s-\frac{\partial g_s^{ij}}{\partial t_l}\pi_{i}\pi_{j}\delta t_l-\frac{\partial g_s^{ij}}{\partial t_s}\pi_{i}\pi_{j}\delta t_s\right.\nn\\
&-&\left.\frac{1}{2}g_s^{ij}\frac{\partial p_i}{\partial t_l}\pi_{i}\pi_{j}\delta t_l-\frac{1}{2}g_s^{ij}\frac{\partial p_i}{\partial t_s}\pi_{i}\pi_{j}\delta t_l-\frac{1}{2}g_s^{ij}\frac{\partial p_j}{\partial t_l}\pi_{i}\pi_{j}\delta t_l-\frac{1}{2}g_s^{ij}\frac{\partial p_j}{\partial t_s}\pi_{i}\pi_{j}\delta t_l\right)\frac{dt_s}{ds}-g_s^{ij}\pi_{i}\pi_{j}\frac{d\delta t_s}{ds}\bigg]\bigg\}ds\nn\;.
\end{eqnarray}
%where $\dot{x} \equiv \frac{\partial x}{\partial t_l}$ and $\mbox{\[\accentset{\ast}{x}\]}\equiv \frac{\partial x}{\partial t_s}$. 
Integrating by parts the first and last terms inside each square bracket, the cancellation among the terms will give
\begin{eqnarray}
\delta S &=& \int\bigg\{\bigg[\frac{1}{2}\left(\frac{\partial p_i}{\partial t_l}\frac{\partial q^i}{\partial t_s}-\frac{\partial p_i}{\partial t_s}\frac{\partial q^i}{\partial t_l}\right)+\left(\frac{\partial g_l^{ij}}{\partial t_s}-\frac{\partial g_s^{ij}}{\partial t_l}\right)\pi_{i}\pi_{j}+\frac{1}{2}\left(g_l^{ij}\frac{\partial p_i}{\partial t_s}-g_s^{ij}\frac{\partial p_i}{\partial t_l}\right)\pi_{i}\pi_{j}\nn\\
&&+\frac{1}{2}\left(g_l^{ij}\frac{\partial p_j}{\partial t_s}-g_s^{ij}\frac{\partial p_j}{\partial t_l}\right)\pi_{i}\pi_{j}\bigg]\frac{dt_s}{ds}\delta t_l+\bigg[\frac{1}{2}\left(\frac{\partial p_i}{\partial t_s}\frac{\partial q^i}{\partial t_l}-\frac{\partial p_i}{\partial t_l}\frac{\partial q^i}{\partial t_s}\right)+\left(\frac{\partial g_l^{ij}}{\partial t_s}-\frac{\partial g_s^{ij}}{\partial t_l}\right)\pi_{i}\pi_{j}\nn\\
&&+\frac{1}{2}\left(g_l^{ij}\frac{\partial p_i}{\partial t_s}-g_s^{ij}\frac{\partial p_i}{\partial t_l}\right)\pi_{i}\pi_{j}+\frac{1}{2}\left(g_l^{ij}\frac{\partial p_j}{\partial t_s}-g_s^{ij}\frac{\partial p_j}{\partial t_l}\right)\pi_{i}\pi_{j}\bigg]\frac{dt_l}{ds}\delta t_s\bigg\}\;.\nn
\end{eqnarray}
Using the equations of motion,
\begin{eqnarray}
&&\frac{dq^k}{dt_l} = g_l^{ik}\pi_{i}\pi_{k}\;,\quad
\frac{dp_k}{dt_l} =-\frac{\partial g_l^{ij}}{\partial q^k}\pi_{i}\pi_{j}\;,\nn\\
&&\frac{dq^k}{dt_s} = g_s^{ik}\pi_{i}\pi_{k}\;,\quad
\frac{dp_k}{dt_s} = -\frac{\partial g_s^{ij}}{\partial q^k}\pi_{i}\pi_{j}\;,\nn
\end{eqnarray}
we obtain
\begin{eqnarray}
\sum_{i,j,n,k = 1}^{N}\left(\frac{\partial g_l^{ij}}{\partial q^k}g_s^{nk}-\frac{\partial g_s^{ij}}{\partial q^k}g_l^{nk}\right)\pi_{i}\pi_{j}\pi_{n}\pi_{k} = 0\;,\label{eq23}
\end{eqnarray}
which is actually identical to \eqref{eq2}. One may find that it is straightforward to show that this condition holds true for every pair of metric tensors in the case of $N$ degrees of freedom. 
\\
\\
\textbf{Proposition}: For integrable pseudo-geodesic Hamiltonian systems, the following identity 
\begin{eqnarray}
\sum_{i,j,n,k = 1}^{N}\left(\frac{\partial g_l^{ij}}{\partial q^k}g_s^{nk}-\frac{\partial g_s^{ij}}{\partial q^k}g_l^{nk}\right)\pi_{i}\pi_{j}\pi_{n}\pi_{k} = 0\;\label{GMT}
\end{eqnarray}
holds true on solutions of the Hamilton's equations.
%\end{theorem}
%\begin{proof}
\\
\\
Above statement can be verified with explicit computation. Next, we will give direct computation on the compatibility between $g_1$ and $g_2$, the metric tensors associated with the first two Hamiltonians in the hierarchy, for the rational and hyperbolic Calogero's GF systems.
\\
\\
\emph{The rational case}: For simplicity, we consider first the case of three particles. We found that, in rational case, the whole inside the bracket of \eqref{GMT} vanish naturally independent of others under the summation. Therefore, the general case of $N$ particles can be proved as follows. Calculating the term inside the bracket, we obtain
\begin{eqnarray}
%\frac{\partial g_1^{ij}}{\partial q^k}g_2^{nk} &=& -\frac{1}{\prod\limits_{a\neq i}^{N}(q^i-q^a)^{2}}\frac{\partial}{\partial q^k}\bigg[\prod\limits_{a\neq i}^{N}(q^i-q^a)\bigg]\frac{\sum\limits_{b\neq n}^{N}q^b}{\prod\limits_{a\neq n}^{N}(q^n-q^a)}\nn\\
\frac{\partial g_1^{ij}}{\partial q^k}g_2^{nk} &=& -\frac{\sum\limits_{l\neq i}^{N}\bigg[\bigg( \prod\limits_{i\neq a \neq l}^N(q^i - q^a) \bigg)(\delta_{ik}- \delta_{lk})\bigg]\sum\limits_{b\neq n}^{N}q^b}{\prod\limits_{a\neq i}^{N}(q^i-q^a)^{2}\prod\limits_{a\neq n}^{N}(q^n-q^a)}\;,\label{eq3}
\end{eqnarray}
\\
and
\begin{eqnarray}
%\frac{\partial g_2^{ij}}{\partial q^k}g_1^{nk} &=& \frac{\prod\limits_{a\neq i}^{N}(q^i-q^a)\frac{\partial}{\partial q^k}(\sum\limits_{b\neq i}^{N}q^b)-(\sum\limits_{b\neq i}^{N}q^b)\frac{\partial}{\partial q^k}\bigg[\prod\limits_{a\neq i}^{N}(q^i-q^a)\bigg]}{\prod\limits_{a\neq i}^{N}(q^i-q^a)^{2}\prod\limits_{a\neq n}^{N}(q^n-q^a)}\nn\\
\frac{\partial g_2^{ij}}{\partial q^k}g_1^{nk} &=& \frac{\prod\limits_{a\neq i}^{N}(q^i-q^a)(\sum\limits_{b\neq i}^{N}\delta_{bk})-\sum\limits_{b\neq i}^{N}q^b\bigg\{ \sum\limits_{l\neq i}^{N}\bigg[\bigg( \prod\limits_{i\neq a \neq l}^N(q^i - q^a) \bigg)(\delta_{ik}- \delta_{lk})\bigg] \bigg\}}{\prod\limits_{a\neq i}^{N}(q^i-q^a)^{2}\prod\limits_{a\neq n}^{N}(q^n-q^a)}\;. \label{eq4}
\end{eqnarray}
Here, we have suppressed the initial condition of every summation and product since they are all starting from one. We observe that (\ref{eq3}) and (\ref{eq4}) are different by just the term that contains $\delta_{bk}$. So, we divide our proof into two parts. 
%The first part is for the case of $i=k$ and the second one is for $i\neq k$. 
\\\\
Part 1 for the case $i=k$: Consider (\ref{eq4}), the condition for the summation on $b$ becomes $b\neq k$ which means that the kronecker delta functions $\delta_{bk}$ are all zero. Therefore, changing every $i$ appearing in (\ref{eq4}) into $k$ and, since $n$ is always equal to $k$ for goldfish models, we find that (\ref{eq3}) and (\ref{eq4}) are exactly the same. 
\\
\\
Part 2 for the case $i\neq k$: $\delta_{ik}$ always vanish and the kronecker delta functions $\delta_{lk}$ will be one only the term that $l=k$. Also, the kronecker delta functions $\delta_{bk}$ can only be one since there is only one term where $b=k$. Then, (\ref{eq4}) is reduced to 
\begin{eqnarray}
\frac{\partial g_2^{ij}}{\partial q^k}g_1^{nk} &=&\frac{(q^i-q^k)\prod\limits_{i\neq a\neq k}^{N}(q^i-q^a)+(q^k+\sum\limits_{i\neq b\neq k}^{N}q^b)\prod\limits_{i\neq a\neq k}^{N}(q^i-q^a)}{\prod\limits_{a\neq i}^{N}(q^i-q^a)^{2}\prod\limits_{a\neq n}^{N}(q^n-q^a)}\nn\\
&=&\frac{(q^i+\sum\limits_{i\neq b\neq k}^{N}q^b)\prod\limits_{i\neq a\neq k}^{N}(q^i-q^a)}{\prod\limits_{a\neq i}^{N}(q^i-q^a)^{2}\prod\limits_{a\neq n}^{N}(q^n-q^a)}\nn\\
&=&\frac{(\sum\limits_{b\neq k}^{N}q^b)\prod\limits_{i\neq a\neq k}^{N}(q^i-q^a)}{\prod\limits_{a\neq i}^{N}(q^i-q^a)^{2}\prod\limits_{a\neq n}^{N}(q^n-q^a)}\;.\label{CC1}
\end{eqnarray}
\\
Also, (\ref{eq3}) becomes
\begin{eqnarray}
\frac{\partial g_1^{ij}}{\partial q^k}g_2^{nk} &=&\frac{(\sum\limits_{b\neq k}^{N}q^b)\prod\limits_{i\neq a\neq k}^{N}(q^i-q^a)}{\prod\limits_{a\neq i}^{N}(q^i-q^a)^{2}\prod\limits_{a\neq n}^{N}(q^n-q^a)}\;.\label{CC2}
\end{eqnarray}
The equation \eqref{CC1} and \eqref{CC2} are identical. Therefore, this completes the verification of the geodesic compatibility for rational case. 
\\
\\
\emph{The hyperbolic case}: We start the computation in the same fashion as the rational case. Substituting the metric tensors into \eqref{GMT}, we get 
\begin{eqnarray}
%\frac{\partial g_1^{ij}}{\partial q^k}g_2^{nk} &=& -\frac{1}{\prod\limits_{a\neq i}^{N}\sinh(q^i-q^a)^{2}}\frac{\partial}{\partial q^k}\bigg[\prod\limits_{a\neq i}^{N}\sinh(q^i-q^a)\bigg]\frac{\sum\limits_{b\neq n}^{N}e^{-2q^b}}{\prod\limits_{a\neq n}^{N}]\sinh(q^n-q^a)}\nn\\
\frac{\partial g_1^{ij}}{\partial q^k}g_2^{nk} &=& -\frac{\sum\limits_{l\neq i}^{N}\bigg[\bigg( \prod\limits_{i\neq a \neq l}^N\sinh(q^i - q^a) \bigg)\cosh(q^i-q^l)(\delta_{ik}- \delta_{lk})\bigg]\sum\limits_{b\neq n}^{N}e^{-2q^b}}{\prod\limits_{a\neq i}^{N}\sinh(q^i-q^a)^{2}\prod\limits_{a\neq n}^{N}\sinh(q^n-q^a)}\;.\label{eq5}
\end{eqnarray}\\
and\\
\begin{eqnarray}
%\frac{\partial g_2^{ij}}{\partial q^k}g_1^{nk} &=& \dfrac{\prod\limits_{a\neq i}^{N}\sinh(q^i-q^a)\frac{\partial}{\partial q^k}(\sum\limits_{b\neq i}^{N}e^{-2q^b})-(\sum\limits_{b\neq i}^{N}e^{-2q^b})\frac{\partial}{\partial q^k}\bigg[\prod\limits_{a\neq i}^{N}\sinh(q^i-q^a)\bigg]}{\prod\limits_{a\neq i}^{N}\sinh(q^i-q^a)^{2}\prod\limits_{a\neq n}^{N}\sinh(q^n-q^a)}\nn\\
\frac{\partial g_2^{ij}}{\partial q^k}g_1^{nk} &=& \frac{\prod\limits_{a\neq i}^{N}\sinh(q^i-q^a)(-2\sum\limits_{b\neq i}^{N}e^{-2q^b}\delta_{bk})-\sum\limits_{b\neq i}^{N}e^{-2q^b}\bigg\{ \sum\limits_{l\neq i}^{N}\bigg[\bigg( \prod\limits_{i\neq a \neq l}^N\sinh(q^i - q^a) \bigg)\cosh(q^i-q^l)(\delta_{ik}- \delta_{lk})\bigg] \bigg\}}{\prod\limits_{a\neq i}^{N}\sinh(q^i-q^a)^{2}\prod\limits_{a\neq n}^{N}\sinh(q^n-q^a)}\;. \label{eq6}
\end{eqnarray}
Again, for the case of $i=k$, it can be easily seen that both \eqref{eq5} and \eqref{eq6} are identical as we do have in the rational type. 
\\
\\
For the case of $i\neq k$, we start with an observation that, for a pair $(11-22)$, we have
\begin{eqnarray}
& &\left(\frac{\partial g_1^{11}}{\partial q^2}g_2^{22}-\frac{\partial g_2^{11}}{\partial q^2}g_1^{22}\right)\pi_{1}\pi_{1}\pi_{2}\pi_{2} + \left(\frac{\partial g_1^{22}}{\partial q^1}g_2^{11}-\frac{\partial g_2^{22}}{\partial q^1}g_1^{11}\right)\pi_{2}\pi_{2}\pi_{1}\pi_{1}\nn \\
&=& \bigg[\left(\frac{\partial g_1^{11}}{\partial q^2}g_2^{22}-\frac{\partial g_2^{11}}{\partial q^2}g_1^{22}\right)+\left(\frac{\partial g_1^{22}}{\partial q^1}g_2^{11}-\frac{\partial g_2^{22}}{\partial q^1}g_1^{11}\right)\bigg]\pi_{1}\pi_{1}\pi_{2}\pi_{2} = 0 
\end{eqnarray}
and this also holds true for other pairs, i.e., $(11-33)$, $(22-44)$, and so on. Then, simplifying (\ref{eq5}) and (\ref{eq6}) as in the rational case and subtracting them, we obtain
\begin{eqnarray}
\frac{\partial g_1^{ij}}{\partial q^k}g_2^{nk}-\frac{\partial g_2^{ij}}{\partial q^k}g_1^{nk} &=& \frac{2e^{-2q^k}\prod\limits_{a\neq i}^{N}\sinh(q^i-q^a)+\prod\limits_{i\neq a \neq k}^N\sinh(q^i - q^a)\cosh(q^i-q^k)\bigg(\sum\limits_{b\neq k}^{N}e^{-2q^b}-\sum\limits_{b\neq i}^{N}e^{-2q^b}\bigg)}{\prod\limits_{a\neq i}^{N}\sinh(q^i-q^a)^{2}\prod\limits_{a\neq k}^{N}\sinh(q^k-q^a)} \label{eq8}
\end{eqnarray}
The term in the numerator inside the last bracket can be simplified as
\begin{eqnarray}
\sum\limits_{b\neq k}^{N}e^{-2q^b}-\sum\limits_{b\neq i}^{N}e^{-2q^b} &=& e^{-2q^i} + \sum\limits_{i\neq b\neq k}^{N}e^{-2q^b}-e^{-2q^k} -\sum\limits_{i\neq b\neq k}^{N}e^{-2q^b} = e^{-2q^i} - e^{-2q^k}\;.\nn
\end{eqnarray}
Then, (\ref{eq8}) becomes
\begin{eqnarray}
\frac{\partial g_1^{ij}}{\partial q^k}g_2^{nk}-\frac{\partial g_2^{ij}}{\partial q^k}g_1^{nk} &=& \frac{\bigg(2e^{-2q^k}\sinh(q^i-q^k)+(e^{-2q^i} - e^{-2q^k})\cosh(q^i-q^k)\bigg)\prod\limits_{i\neq a\neq k}^{N}\sinh(q^i-q^a)}{\prod\limits_{a\neq i}^{N}\sinh(q^i-q^a)^{2}\prod\limits_{a\neq k}^{N}\sinh(q^k-q^a)}\nn\\
&=& \frac{\bigg(2e^{-2q^k}\sinh(q^i-q^k)-2\sinh(q^i-q^k)\cosh(q^i-q^k)\bigg)\prod\limits_{i\neq a\neq k}^{N}\sinh(q^i-q^a)}{\prod\limits_{a\neq i}^{N}\sinh(q^i-q^a)^{2}\prod\limits_{a\neq k}^{N}\sinh(q^k-q^a)}\nn\\
&=& \frac{\bigg(2e^{-2q^k}-2\cosh(q^i-q^k)\bigg)\prod\limits_{a\neq i}^{N}\sinh(q^i-q^a)}{\prod\limits_{a\neq i}^{N}\sinh(q^i-q^a)^{2}\prod\limits_{a\neq k}^{N}\sinh(q^k-q^a)}\nn\\
 &=& \frac{\bigg(2e^{-2q^k}-2\cosh(q^i-q^k)\bigg)}{\prod\limits_{a\neq i}^{N}\sinh(q^i-q^a)\prod\limits_{a\neq k}^{N}\sinh(q^k-q^a)}\;.\label{eq9}
\end{eqnarray}
The summations appear in (\ref{eq2}) will generate another term similar to (\ref{eq9}) but the indices are interchanged such that (\ref{eq9}) becomes
\begin{eqnarray}
\frac{\partial g_1^{nk}}{\partial q^i}g_2^{ij}-\frac{\partial g_2^{nk}}{\partial q^i}g_1^{ij} &=& \frac{\bigg(2e^{-2q^i}-2\cosh(q^k-q^i)\bigg)}{\prod\limits_{a\neq k}^{N}\sinh(q^k-q^a)\prod\limits_{a\neq i}^{N}\sinh(q^i-q^a)}\;.\label{eq21}
\end{eqnarray}
Since the cosine hyperbolic is an even function, and the denominators are the same, adding (\ref{eq9}) and (\ref{eq21}) up as suggested by the observation, we get
\begin{eqnarray}
(2e^{-2q^k}-2\cosh(q^i-q^k))+ (2e^{-2q^i}-2\cosh(q^k-q^i)) = 2(e^{-2q^k}+e^{-2q^i})-4\cosh(q^k-q^i) = 0\nn\;.
\end{eqnarray}
This completes the verification for the hyperbolic case.

\section{Geometrical interpretation} \label{section4}
\begin{figure}[h]
 	\centering
 	\includegraphics[width=0.7\linewidth]{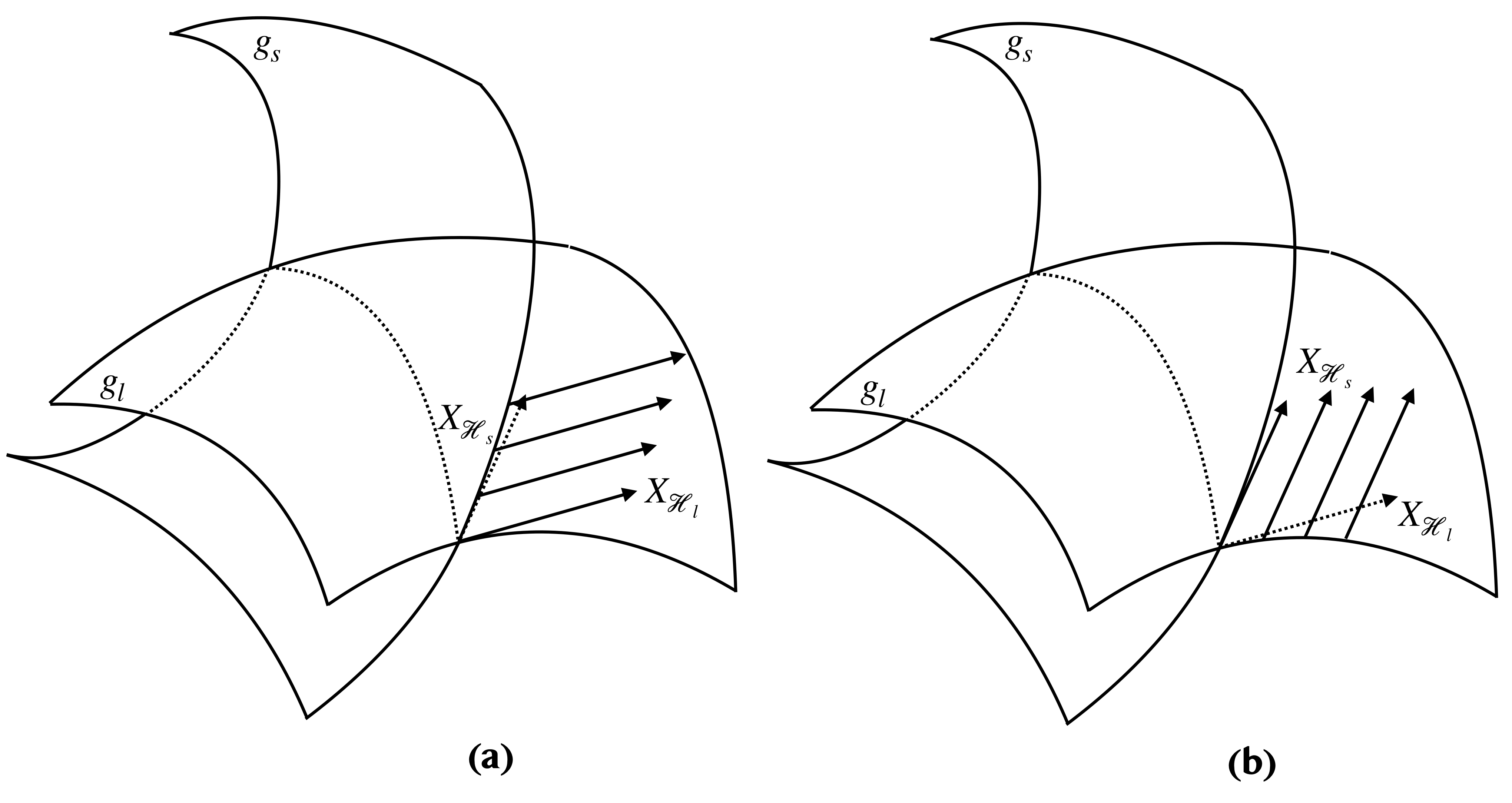}
 	\caption{Geodesic compatibility: (a) parallel transport of the vector field $X_{\mathscr{H}_l}$ in the direction of the $X_{\mathscr{H}_s}$. (b) parallel transport of the vector field $X_{\mathscr{H}_s}$ in the direction of the $X_{\mathscr{H}_l}$. Here, the different Hamiltonian flows are represented by 2-dimensional sheet governed by different metric tensors.}
 	\label{GEO11}
 \end{figure}

To see how we would interpret the geodesic compatibility in terms of the geometry, the present form of the relation \eqref{GMT} is not suitable since the Hamiltonians \eqref{eq11} are not in the canonical variable representation. We then assume that there exists an integrable system with the Hamiltonian hierarchy 
\begin{equation}
\{\mathscr{H}_l(p,q)=g^{ij}p_ip_j\;;\; l=1,2,3,...,N\}\;,\label{Hhe}
\end{equation}
where $p_i$ are the canonical momenta, and $g^{ij}=g^{ij}(q)$ are the metric tensors as functions of canonical coordinates $q=\{q^1,q^2,...,q^N\}$. The involution feature (Hamiltonian commuting flows) between two Hamiltonians,
\begin{eqnarray}
    \mathscr{H}_{l} = g_l^{ij}p_{i}p_{j} \quad \mbox{and} \quad \mathscr{H}_{s} = g_s^{ij}p_{i}p_{j}\;,
\end{eqnarray}
gives
\begin{eqnarray}
    \sum_{i,j,n,k = 1}^{N}\left(\frac{\partial g_l^{ij}}{\partial q^k}g_s^{nk}-\frac{\partial g_s^{ij}}{\partial q^k}g_l^{nk}\right)p_{i}p_{j}p_{n} = 0. \label{GMT1}
\end{eqnarray}
which is a geodesic compatibility condition. To unravel the geometrical insight, we employ the relation between the derivative with respect to coordinates $q^i$ and the affine (Levi-Civita) connection
\begin{eqnarray}
    \partial_{k}g^{ij} = -\Gamma^{i}_{kh}g^{hj}-\Gamma^{j}_{kh}g^{ih}\;.
\end{eqnarray}
Then, we find that the term inside the bracket of the geodesic compatibility condition \eqref{GMT1} becomes
\begin{eqnarray}
    \partial_{k}g^{ij}_l g^{nk}_s-\partial_{k}g^{ij}_s g^{nk}_l&=& (-\Gamma^{i}_{kh}g^{hj}_l -\Gamma^{j}_{kh}g^{ih}_l)g^{nk}_s - (-\Gamma^{i}_{kh}g^{hj}_s -\Gamma^{j}_{kh}g^{ih}_s)g^{nk}_l\;.\label{comaff}
\end{eqnarray}
The geodesic compatibility can now be written as
\begin{eqnarray}
&&\sum_{i,j,n,k = 1}^{N}\left(\frac{\partial g_l^{ij}}{\partial q^k}g_s^{nk}-\frac{\partial g_s^{ij}}{\partial q^k}g_l^{nk}\right)p_{i}p_{j}p_{n} \nn\\&=& -\sum_{i,j,n,k = 1}^{N}\bigg[(\Gamma^{i}_{kh}g^{hj}_l +\Gamma^{j}_{kh}g^{ih}_l)g^{nk}_s - (\Gamma^{i}_{kh}g^{hj}_s +\Gamma^{j}_{kh}g^{ih}_s)g^{nk}_l\bigg]p_{i}p_{j}p_{n}=0\;.\label{GMT2}
\end{eqnarray}
Recalling the covariant derivative of the metric tensor
\begin{eqnarray}
    \nabla_{k}g^{ij} = \partial_{k}g^{ij} +\Gamma^{i}_{kh}g^{hj}+\Gamma^{j}_{kh}g^{ih}\;, \label{nabla}
\end{eqnarray}
and substituting (\ref{nabla}) into the term inside the bracket of \eqref{GMT1}, we get
\begin{eqnarray}
    \partial_{k}g^{ij}_l g^{nk}_s-\partial_{k}g^{ij}_s g^{nk}_l&=& (\nabla_{k}g^{ij}_l-\Gamma^{i}_{kh}g^{hj}_l-\Gamma^{j}_{kh}g^{ih}_l)g^{nk}_s - (\nabla_{k}g^{ij}_s-\Gamma^{i}_{kh}g^{hj}_s-\Gamma^{j}_{kh}g^{ih}_s)g^{nk}_l\nn\\
    &=& [(\nabla_{k}g^{ij}_l)g^{nk}_s-(\nabla_{k}g^{ij}_s)g^{nk}_l] \nn\\ &&- [(\Gamma^{i}_{kh}g^{hj}_l +\Gamma^{j}_{kh}g^{ih}_l)g^{nk}_s - (\Gamma^{i}_{kh}g^{hj}_s +\Gamma^{j}_{kh}g^{ih}_s)g^{nk}_l]\;.\label{com23}
\end{eqnarray}
%Consider four terms outside the square bracket. Interchanging $i$ and $j$ in the second and fourth terms, we have
%\begin{eqnarray}
%    \partial_{k}g^{ij}_l g^{nk}_s-\partial_{k}g^{ij}_s g^{nk}_l&=&  [(\nabla_{k}g^{ij}_l)g^{nk}_s-(\nabla_{k}g^{ij}_s)g^{nk}_l] \nn\\ &&+ g^{hj}_l\Gamma^{h}_{ik}g^{nk}_s + g^{hj}_l\Gamma^{h}_{ik}g^{nk}_s -g^{hj}_s\Gamma^{h}_{ik}g^{nk}_l - g^{hj}_s\Gamma^{h}_{ik}g^{nk}_l\nn\\
  %  &=& [(\nabla_{k}g^{ij}_l)g^{nk}_s-(\nabla_{k}g^{ij}_s)g^{nk}_l]+ %2\Gamma^{h}_{ik} (g^{hj}_lg^{nk}_s - g^{hj}_sg^{nk})
%\end{eqnarray}
The second bracket in \eqref{com23} effectively vanishes according to \eqref{GMT2} resulting in
\begin{eqnarray}
   \sum_{i,j,n,k = 1}^{N}\left(\frac{\partial g_l^{ij}}{\partial q^k}g_s^{nk}-\frac{\partial g_s^{ij}}{\partial q^k}g_l^{nk}\right)p_{i}p_{j}p_{n}=\sum_{i,j,n,k = 1}^{N}\left( (\nabla_{k}g^{ij}_l)g^{nk}_s-(\nabla_{k}g^{ij}_s)g^{nk}_l\right)p_{i}p_{j}p_{n}=0 \;.\label{comnabla}
\end{eqnarray}
We know that \eqref{comnabla} is a direct consequence of the Hamiltonian commuting flows, see also section \ref{section3}, $\{ \mathscr {H}_l,\mathscr{H}_s\}$=0. Suppose that $X_{\mathscr {H}_l}$ and $X_{\mathscr {H}_s}$ are vector fields associated with the Hamiltonians $\mathscr {H}_l$ and $\mathscr {H}_s$, respectively. We have now a condition that the Lie bracket of these two vector fields vanishes,
\begin{equation}
\left[ X_{\mathscr {H}_l},X_{\mathscr {H}_s}\right]=0\;.\label{LieB}
\end{equation}
It is well-known that \eqref{LieB} gives a compatibility between two Hamiltonian vector fields. This means that compatibility between flows forms a perfect parallelogram. The covariant derivative \eqref{nabla} gives the parallel transport of the vector on the manifold. Then, \eqref{comnabla} may be treated as the compatible parallel transports of two different Hamiltonian vector fields, see figure \ref{GEO11}.
%The right-hand side of (\ref{comnabla}) can be physically interpreted as a measure of the compatibility between the parallel transport of a vector which belongs to a vector field on a Riemannian manifold with metric $g^{ij}_l$ along the path on another manifold with metric $g^{nk}_s$ and the parallel transport of another vector on a Riemannian manifold with metric $g^{ij}_s$ along the path on a manifold with metric $g^{nk}_l$. This interpretation also leads to a close relationship between this condition and the Lie bracket. In the case of rational GF model, (\ref{comnabla}) vanishes which implies that different orders of parallel transports between two different Riemannian manifolds are compatible with each other. However, for the hyperbolic case, it appears that the right-hand side of (\ref{comnabla}) does not vanish individually like in the rational case. The momentum terms are necessary. Therefore, this means that hyperbolic GF model possesses a non-trivial affine connection.
%
%
%
\section{Conclusion}\label{section5}
We have successfully constructed the geodesic compatibility condition to capture the multidimensional consistency in terms of metric tensors. This compatibility between metric tensors, which is equivalent to the Hamiltonian commuting flows and the closure relation, can possibly be treated as an integrability feature of the system. The rational and hyperbolic GF systems, hierarchy geodesic flows \eqref{eq11}, are used as concrete examples to explicitly verify the geodesic compatibility condition. The condition can be geometrically interpreted as  compatible parallel transports between two different directions corresponding to two Hamiltonian vector fields. We put here a remark on the RS type systems. The geodesic interpretation holds only for the first flow in the RS hierarchy since the second equation of motion in the hierarchy is not in the geodesic form \cite{Sikarin2}. Then, the RS type systems is not applicable for the geodesic compatibility test. Another point is that we do not have the geodesic interpretation in the Lagrangian description for both GF and RS systems, see the Lagrangian hierarchy in \cite{Sikarin2, Umpon}.

\end{document}